\documentclass[reprint,superscriptaddress,nofootinbib,amsmath,amssymb,aps,prl,onecolumn,notitlepage]{revtex4-1}

\usepackage[a4paper, margin=2.50cm]{geometry}

\usepackage{bbm}
\usepackage{graphicx}
\usepackage{fancyhdr, color}
\usepackage{hyperref} 
\usepackage{amsmath}
\usepackage{amssymb}
\usepackage{graphicx}
\usepackage{color}
\usepackage{subfigure}
\usepackage{ulem}
\usepackage{enumitem} 
\usepackage{bm}
\newcommand{\bra}[1]{\langle#1|}
\newcommand{\ket}[1]{|#1\rangle}

\begin{document}

\fancyhead[C]{\sc \color[rgb]{0.4,0.2,0.9}{Quantum Thermodynamics book}}
\fancyhead[R]{}

\title{Quantum Thermometry}

\author{Antonella De Pasquale}
\email{adepasquale83@gmail.com} 
\affiliation{Department of Physics and Astronomy, University of Florence, Via G. Sansone 1, 50019, Sesto Fiorentino (FI), Italy}
\affiliation{INFN Sezione di Firenze, via G.Sansone 1, I-50019 Sesto Fiorentino (FI), Italy}
\affiliation{NEST, Scuola Normale Superiore and Istituto Nanoscienze-CNR, I-56126 Pisa, Italy}
\author{Thomas M. Stace}
\affiliation{ARC Centre for Engineered Quantum System, Department of Physics, University of Queensland, Brisbane, QLD 4072, Australia}

\date{\today}

\begin{abstract}
We discuss the application of techniques of quantum estimation theory and quantum metrology to thermometry. The ultimate limit to the precision at which the temperature of a system at thermal equilibrium can be determined is related to the heat capacity when global measurements are performed on the system. We prove that if technical or practical limitations restrict our capabilities to local probing, the highest achievable accuracy to temperature estimation reduces to a sort of mesoscopic version of the heat capacity. Adopting a more practical perspective, we also discuss the relevance of qubit systems as optimal quantum thermometers, in order to retrieve the temperature, or to discriminate between two temperatures, characterizing a thermal reservoir. We show that quantum coherence and entanglement in a probe system can facilitate faster, or more accurate measurements of temperature. While not surprising given this has been demonstrated in phase estimation, temperature is not a conventional quantum observable, so that these results extend the theory of parameter estimation to measurement of non-Hamiltonian quantities. Finally we point out the advantages brought by a less standard estimation technique based on sequential measurements, when applied to quantum thermometry. 
\end{abstract}

\maketitle

\thispagestyle{fancy}

\section{Introduction}

When studying a physical system, there are physically important quantities that cannot be directly measured, either in principle or due to some technical obstructions. Temperature represents a paradigmatic example: it is a non-linear function of the density matrix, so it is not a (quantum) observable of the system. Instead, to infer its value, we need to measure another physical quantity, such as the mean kinetic energy, which is related to the quantity of interest.   The theory of estimation provides the formal framework in order to tackle these kinds of indirect measurements. 

This situation is precisely the one faced in modern primary and secondary thermometry standards.  For instance, current primary thermometers are based on the resonant frequency of a precisely machined microwave resonator filled with a noble gas \cite{mohr}.  The temperature dependence of the refractive index of the gas is theoretically calculable to high accuracy, and so the resonant frequency of the system provides an indirect measurement of the temperature, via the geometry and the refractive index of the gas-filled resonator. In the case of ultra-high precision secondary thermometers \cite{weng:2014}, the same operational principle applies: the resonant frequency of a toroidal micro-resonator depends on the  refractive index of the glass medium, which is itself a thermodynamic (and thus temperature-dependent) quantity, albeit in practice too complicated to calculate at sufficiently high accuracy from first principles. Moreover, within the plethora of recent theoretical efforts aiming at a self-consistent generalization of the classical thermodynamics to small-scale physics, where quantum effects become predominant~\cite{Allahverdyan2002, Hilt09, Williams11}, precision nanothermometry  proved to be quite successful in exploiting quantum effects~\cite{Brunelli11,Marzolino13,Salvatori14,Correa15}. 

Throughout this chapter, after introducing some general tools both of classical and quantum estimation theory, we focus on different thermometric tasks, also discussing the role played by quantum correlations and coherence in order to enhance the accuracy level achievable when measuring the temperature. We start by considering systems of arbitrary dimensions with no restrictions on the structure of the corresponding Hamiltonian. Then we focus on the case of single qubit thermometry. 

\section{Quantum Estimation Theory and Quantum Metrology}

To begin the discussion of the formalism, we start by considering the  reconstruction of an unknown parameter $\lambda$, which we suppose is not directly observable, but which we can connect via theoretical considerations to a $\lambda$-dependent quantity $\Theta$, which can be directly measured.  Once a large sequence of independent identically distributed measurements of $\Theta$, $\vec{\theta}=\{\theta_1, \theta_2, \ldots \theta_N\}$ ($N\gg1$), are collected, the value of $\lambda$ is recovered in the form of a random variable $\lambda^{(\mathrm{est})}$, representing the estimation of $\lambda$ retrieved from $\vec\theta$.

The ultimate precision limit on the estimation of $\lambda$ is given by the Cram\'er-Rao bound~\cite{Cramer46} on the Root Mean Square Error (RMSE) $\Delta \lambda=\sqrt{\mathbb{E}[(\lambda^{(\mathrm{est})} - \lambda)^2]}$:
\begin{equation} \label{eq:CramerBOUND}
  \Delta \lambda \geq \frac{1}{\sqrt{N {\cal F(\lambda)}}} \\
 \end{equation} 
with 
\begin{equation}\label{def:FisherInfo}
 \mathcal{F}(\lambda) =   \int \!d \theta \frac{1}{p(\theta|\lambda)} \left(\frac{\partial  p(\theta|\lambda)}{\partial \lambda}\right)^2\,.
\end{equation} 

Here $\mathbb{E}[x]$ indicates the expectation value of the random variable $x$, and $p(\theta|\lambda)$ is the conditional probability of measuring $\theta$ (here assumed continuous for simplicity) if the value of the parameter under consideration is $\lambda$.
The quantity $\mathcal{F}$ is the \textit{Fisher Information} (FI), and derives from the Fisher-Rao distance between probability distributions differing by an infinitesimal increment in $\lambda$, namely $p(\theta|\lambda)$ and $p(\theta|\lambda + \delta \lambda)$. 

The Cram\'er-Rao bound basically sets the rules for recognizing whether the procedure followed in order to retrieve $\lambda$ is optimal or not. 
On the one hand, the probability distribution $p(\theta|\lambda)$ (and thus its sensitivity to small variations of the parameter of interest $\lambda$) depends on the chosen measurement: optimal measurements are those with conditional probability maximizing the Fisher Information.
On the other hand, for any fixed measurement, an efficient estimator is the one that saturates the Cramer-Rao inequality. If the data sample is sufficiently large, it results that an efficient estimator is provided by the maximum-likelihood principle, based on the intuition that the observed data $\vec{\theta}$ have been measured since they hold the highest probability to be obtained
\footnote{
The maximum-likelihood principle selects the parameter values that make the data most probable. It stems from the definition of the likelihood function ${\cal L}(\lambda)$ as the joint conditional probability of the observed data, that for the case of independent measurements reduces to the product of the probabilities of the single outcomes $\theta_i$, 
\begin{equation}
 {\cal L}(\lambda)={\cal L}(\vec{\theta}|\lambda)=\prod_{i}^{N} p(\theta_i|\lambda).
\end{equation}
The maximum-likelihood estimate is the value of $\lambda$ that maximizes ${\cal L}(\lambda)$ or equivalently its logarithm. This procedure selects the parameter values that make the data most probable. It results that the variance on the maximum-likelihood estimate of $\lambda$, in the limit of large $N$, saturates the 
Cram\'er-Rao bound~(\ref{eq:CramerBOUND}).
}.
This is a typical example of an asymptotically efficient estimator. There also exist special families of probability distributions allowing for the construction of an estimator with only a finite number of measurements.

The pre-factor $1/\sqrt{N}$ in the Cram\'er-Rao bound~(\ref{eq:CramerBOUND}) is due to the additivity of the Fisher Information for the case of independent measures, and is a direct consequence of the central limit theorem according to which the average of a large number $N$ of independent measurements (each having a standard deviation $\Delta \sigma$) converges to a Gaussian distribution with standard deviation $\Delta \sigma/\sqrt{N}$, thus yielding the scaling $1/\sqrt{N}$ on the error on the average. We will return to this point  when, in the context of single qubit thermometers, we will compare the estimation strategy on independent measurements, with a protocol based on measurements of multi-particle correlated states, see Equations~(\ref{CR})-(\ref{CRSMS}).

\begin{figure}[t]
	\begin{center}
	{\includegraphics[width = 1 \textwidth]{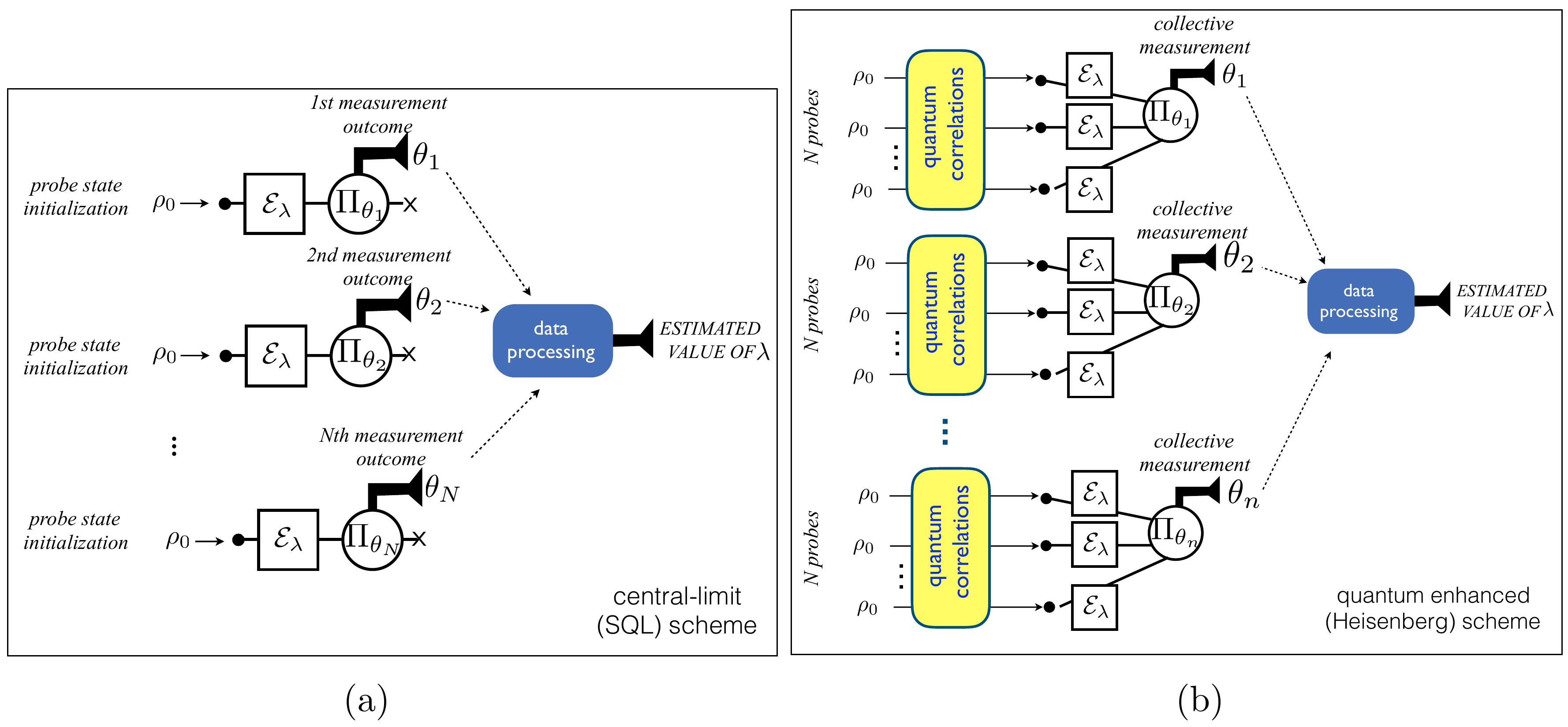}}
\caption{\label{fig:schemes} Schematic representation of typical estimation schemes. The central-limit scheme, in panel (a), refers to the preparation of $N$ independent probes prepared in the same state and separately measured, yielding a precision in the estimation of $\lambda$ scaling as $1/\sqrt{N}$. In panel (b) is shown a schematic representation of an estimation protocol based on the introduction of quantum correlations in the probe preparation stage and eventually on nonlocal measurements, leading to the Heisenberg bound $1/(N\sqrt{n})$ in the estimation of $\lambda$, where $N$ is the number of probes in each entangled block, and $n$ is the number of repeated measurements.}
\end{center}
\end{figure}

In high-precision measurements, and in the quantum regime, it is important to include a proper accounting of the measurement process and apparatus. We introduce an ancillary physical system, the probe,  over which we assume a high degree of control.  As the system of interest interacts with the probe, information about the system quantity $\lambda$ is encoded into the state of the probe. We then make a measurement of the probe, in order to make an inference about $\lambda$.  This protocol is summarised as follows:
\begin{enumerate}[label=\textit{\roman*)}]
\item Probe initialization: the probing system is prepared in an assigned state  $\rho_0$.
\item Probe evolution: the probe interacts with the system, and evolves according to a $\lambda$-dependent process described by a superoperator $\mathcal{E}_\lambda$, so as to imprint $\lambda$ onto the probe state, via $\rho_\lambda =\mathcal{E}_\lambda (\rho_0) $. 
\item Probe readout:  a (quantum) measurement is performed on $\rho_\lambda$,  followed by classical data processing on the outcomes. This is what is properly defined as the `estimation step'.   
\end{enumerate}
This sequence is repeated  for  $N$ independent probes, all initialized in the same initial state $\rho_0$.  Fig.~\ref{fig:schemes}(a) illustrates this protocol schematically. 

By reference to the practical primary and secondary thermometry examples introduced at the start of this chapter, the `system' consists of the thermodynamic medium (gas or glass), with a temperature dependent refractive index.  The `probe' consists of the electromagnetic modes that couple to the medium.  The measurement step consists of measuring the resonant frequency of the probe, which is both an observable and is practically accessible.

Formally, the general connection between the final state of the probe and the desired measurement result is achieved by expressing the measurement in terms
of a set of positive operators $\Pi_\theta$,  realizing a partition of unity $\int \!d \theta \,\Pi_{\theta} = \mathbb{I}$, that form a Positive Operator Valued Measure (POVM). The conditional probability of obtaining the outcome $\theta$ if the probe state is $\rho_\lambda$ is given by $p(\theta|\lambda)=\mathrm{Tr}[ \Pi_\theta \rho_\lambda]$. We note  that $\theta$ does not necessarily represent the eigenvalue of a quantum observable:  POVMs are more general than projective measurements, including for instance the possibility of measurement imperfections.

Classically, a probe could be designed in principle to encode $\lambda$ with arbitrary precision, in which case measurements of $\lambda$ would be correspondingly precise.
However,  the Heisenberg uncertainty relations constrain the capability of quantum systems.  Furthermore, the finite size of the probe also limits the amount of information it can encode.  The main objective of quantum estimation theory is to optimize the measurement protocol, to obtain the best estimate of $\lambda$ given the resources (probe size, number of probes, interaction time, etc) available. More precisely, it has been shown that the bound in Eq.~(\ref{eq:CramerBOUND}) can be further boosted  by maximizing the FI  with respect to all possible POVMs, yielding
\begin{eqnarray} \label{eq:quantumCramerBOUND}
  \Delta \lambda \geq \frac{1}{\sqrt{N  \; \underset{\{\Pi_\theta\}}{\max}  {\cal F(\lambda)}}} \geq \frac{1}{\sqrt{N \cal Q(\lambda)}}\,.
\end{eqnarray}
The term on the right is known as the quantum Cram\'er-Rao bound and is written in terms of the so-called Quantum Fisher Information (QFI) $\cal Q(\lambda)$~\cite{Braunstein96}. The physical meaning of this functional is rooted in the geometrical structure of the stastical model used to parametrize the Hilbert space of the probe. Indeed ${\cal Q}(\lambda)$ can be expressed as the infinitesimal variation of the probing system quantified by the Bures distance ${\cal D}_B$~\cite{Bures69}, i.e.\
\begin{equation}\label{def:QFI}
{\cal Q}(\lambda) = 4 \lim_{\delta \lambda \to 0 } \frac{{\cal D}_B^2 (\rho_\lambda,\rho_{\lambda+\delta \lambda})}{ \delta \lambda^2}
=8 \lim_{\delta \lambda \to 0 } \frac{1 - F(\rho_\lambda, \rho_{\lambda+ \delta \lambda})}{ \delta \lambda^2}\,,
\end{equation}
where $F(\rho, \rho') = \mathrm{Tr}[ \sqrt{ \sqrt{\rho} \rho'  \sqrt{\rho}}]$ is the Uhlmann fidelity between the states $\rho$ and $\rho'$~\cite{Peres84,Jozsa94}.
The quantum Cram\'er-Rao bound holds for all possible POVMs on the $N$ probes, including joint measurements
that might exploit quantum resources like entanglement~\cite{Giovannetti11}. It is known that the bound in Eq.~(\ref{eq:quantumCramerBOUND}) is achievable through estimation strategies exploiting only local operations and classical communication~\cite{Hayashi05,Hayashi06}. More precisely, a sufficient condition~\cite{Helstrom67,Holevo82} for saturating the quantum Cram\'er-Rao bound is given by the use of a POVM given by one-dimensional projection operators onto the eigenstates of the so-called Symmetric Logarithmic Derivative (SLD) $L_\lambda$, a selfadjoint operator satisfying the equation
\begin{equation}
\frac{\partial \rho_\lambda}{\partial \lambda}=\frac{L_\lambda \rho_\lambda + \rho_\lambda  L_\lambda }{2}\,.
\end{equation}
Indeed it results that the QFI can be also computed as ${\cal Q}(\lambda)=\mathrm{Tr}[\rho_\lambda  L_\lambda^2]$.  It is important to notice that although the use of this optimal POVM is sufficient to saturate the bound~(\ref{eq:quantumCramerBOUND}), such optimal measure depends, in general, on the true value of the parameter one wants to estimate, therefore asking for adaptive estimation strategies~\cite{Fischer00, Fujiwara06}.

The paradigmatic quantum estimation problem in the literature is to estimate the phase $\lambda$ parametrizing a unitary transformation on $\rho_0$, ${\cal E}_\lambda(\rho_0)=e^{-i \lambda H}\rho_0e^{i \lambda H}$ generated by the `Hamiltonian' $H$, which we assume to be independent 
of $\lambda$. In this simple scenario, also the QFI is independent of $\lambda$, and is given by~\cite{Caves94} 
\begin{equation}
 {\cal Q} = 4\sum_{i<j} \frac{(\phi_i - \phi_{j})^2}{\phi_i + \phi_{j}}|\langle \phi_i | H | \phi_j\rangle|^2\,,
  \label{eq:QFI_nodisturb}
\end{equation}
where $\phi_j$ and $|\phi_j\rangle$ are the eigenvalues and the eigenvectors of $\rho_0$, respectively, i.e. \ $\rho_0=\sum_j \phi_j |\phi_j\rangle \langle \phi_j|$. 
From the property of strong concavity for the fidelity
\footnote{Given $p_i$ and $p'_i$ two probability density distributions, and $\{\rho_i \}$ and $\{\rho'_i\}$ two sets of density matrices, it results that ${\cal F}\left( \sum_{i} p_i \rho_i, \sum_{i} p'_i \rho'_i   \right) \geq \sum_{i}\sqrt{p_i p'_i}{\cal F} ( \rho_i, \rho'_i )$. This property is dubbed strong concavity property for the fidelity.
}
${\cal Q}$ is maximised for pure states $\rho_0 = |{\phi_0} \rangle \langle \phi_0|$, and it is proportional to the variance of $H$,
\begin{equation}\label{eq:Qpure}
 {\cal Q}(\lambda) = 4 \left(\langle{\phi_0}| H^2 |\phi_0 \rangle - \langle{\phi_0}| H |\phi_0 \rangle^2\right)\,.
\end{equation} 
Indeed it results that in this case the SLD is simply given by $L_\lambda=2 (H - \langle{\phi_0}| H |\phi_0 \rangle)$.
It follows that the optimal state maximizing the value of ${\cal Q}$ is the equally weighted superposition 
of the eigenvectors corresponding to the maximum, $h_{\max}$, and minimum, $h_{\min}$, eigenvalues of $H$, i.e.\ \mbox{$|\phi_0^{\rm(opt)}\rangle = \frac{1}{\sqrt{2}} (|h_{\max}\rangle+|h_{\min}\rangle)$}, yielding ${\cal Q}(\lambda)=(h_{\max} - h_{\min})^2$. 

Returning to the quantum Cram\'er-Rao bound in equation~(\ref{eq:quantumCramerBOUND}), the central-limit scaling $1/\sqrt{N}$, known colloquially (and somewhat misleadingly) in the literature as the Standard Quantum Limit (SQL), is the benchmark of estimation strategies based on independent identically distributed (i.i.d.) variables.  In optical experiments, the SQL  manifests as a significant technical noise floor (e.g.\ in homodyne or heterodyne field measurements), and is a consequence of quantum shot noise, associated to the Poissonian arrival times of quantised photons in a coherent state. From a mathematical point of view, it is a direct consequence of the additivity of the Quantum Fisher Information when applied to the tensor states  $(\rho_\lambda)^{\otimes N}$, describing the global state of the $N$ independent probes. 

Despite its evocative name, the SQL is not a fundamental limit, and can be broken. 
In particular it has been shown~\cite{Giovannetti04} that in absence of external noise, (e.g.\ for unitary phase estimation), the quantum Cram\'er-Rao bound implies that the use of entanglement between the $N$  probe systems, together with joint measurements on the probes, leads to an improvement in precision by a factor of $\sqrt{N}$. In other words, if we change the measurement protocol to use a total of $\nu=n N$ probes, in which we entangle blocks of $N$ probes, and repeat $n$ times (as shown in Fig.~\ref{fig:schemes}(b)) we get a measurement precision of
\begin{equation}\label{eq:HeisenbegBound}
  \Delta \lambda \geq \frac{1}{N\sqrt{n \cal Q(\lambda)}}\,.
\end{equation}
The improved scaling $\sim1/N$, known as the ``Heisenberg bound'', stems only from the employment of quantum resources in the preparation stage.  The bound in Eq.~(\ref{eq:HeisenbegBound}) may be achievable with adaptive strategies based on a POVM that acts locally on each probe~\cite{Giovannetti06}.

When the number $n$ of repetitions is small,  special instances of phase estimation have  been found in which a scaling of the order of $\nu^{-1}\log(\nu)$ can be achieved~\cite{deBurgh05,Ji08}. Of practical significance is to characterize optimal performances in presence of external disturbance: many discouraging results attest to the fragility of entanglement which, in a noisy environment, limits any precision improvement at most to a constant factor independent of $N$~\cite{Huelga97}, or to a super-classical precision scaling $N^{5/6}$, achieved when the perturbation involves a preferential direction perpendicular to the unitary evolution governed by the parameter to be estimated~\cite{Chaves13}. As yet, an exhaustive answer, providing a systematic method for taking into account the presence of noise, is still missing. 

One of the challenges of quantum metrology is  to  design protocols that achieve Heisenberg-like scaling for various estimation purposes~\cite{Giovannetti11, Kolodynski97, Escher11,DePasquale13}.  The paradigm of phase estimation in quantum optical systems has been an influential motivator~\cite{Caves81,Yurke86, Dowling98, Barnett03}, and has specific applications in gravitational wave detection  and biological microscopy \cite{bowen}. 

\section{Quantum Thermometry}

Let us now apply the tools of quantum estimation theory  to a measuring temperature.  Temperature is not a quantum observable, unlike the phase in an interferometer, so it necessarily must be inferred from an intermediate quantity\footnote{Although we note that the `phase operator' of a harmonic oscillator is itself only properly defined in a limiting sense, much like the position operator of a free particle: one can write a limit of hermitian operators that localise the quantity, but they are accompanied by a divergence in the conjugate variable (number or position), which leads to unphysical energetic divergences.  However, this is different from temperature, which cannot be defined as the limit of a sequence of hermitian operators.}.  This subtlety makes the problem paradigmatic for quantum measurement of more general non-Hamiltonian quantities.

In the examples of high-precision thermometers given above, and in almost all other practical thermometers, the measurement device comes into thermal equilibrium with the bath that is being measured.   We call such a device a  `thermalising' thermometer. 

Consider a  quantum system ${\cal S}$ at thermal equilibrium with an external bath at temperature $T$. The state of $\cal S$ is described by the canonical Gibbs ensemble $\rho_\beta = e^{- \beta H}/{\cal Z}_{\beta}$, where $H$ is the system Hamiltonian, $\beta = 1/(k_B T)$ the inverse temperature of the system with $k_B$ the Boltzmann constant, and ${\cal Z}_{\beta}=\mathrm{Tr}[e^{-\beta H}]$ the associated partition function. Notice that with respect to the three-step scheme of typical quantum estimation protocols discussed earlier, we are assuming to have already carried out the first two duties, $\rho_\beta$ being the state of $\cal S$ already holding a dependence on $T$. Here we will focus on the so-called estimation step.

In \cite{stace:2010a}, a very simple argument is given establishing that if we assume that a thermalising thermometer has an extensive thermodynamic energy, i.e.\ $\bar E\equiv-\frac{\partial \ln \mathcal Z_\beta}{\partial \beta}=N\bar\varepsilon(\beta)$, where $\bar\varepsilon$ is the average internal energy per particle, then the uncertainty (measured by the root-mean-square error) in the measurement of $\beta$ is
\begin{equation}
{\Delta \beta}{}\geq\frac{1}{\sqrt{N}}\frac{1}{\sqrt{\bar\varepsilon'}},\label{eqn:relunc}
\end{equation}
where $\bar \varepsilon'=|d\bar \varepsilon/d\beta|$. We review the argument in \cite{stace:2010a} briefly.  Consider a thermometer in the thermal state
$
\rho=e^{-\beta H}/{{\cal Z}_\beta}$.  Assuming that the average internal energy of the thermometer is extensive, it is given by $\bar E
=N\bar\varepsilon(\beta)$. 
Since $\bar \varepsilon(\beta)$ is a monotonically increasing function of temperature, the sample standard deviation of $\bar \varepsilon$ and $\beta$ are related by the identity
$
\Delta \beta ={\Delta \varepsilon}/{\bar \varepsilon'}
$. 
 To calculate ${\Delta\varepsilon}$, note that the sample variance in the total internal energy is given by 
\begin{equation}
\Delta^2E=\frac{\partial^2 \ln {\cal Z}_\beta}{\partial \beta^2}=-\frac{\partial \bar E}{\partial \beta}=N\bar \varepsilon'\,,
\end{equation}
which demonstrates that the variance in the total internal energy of the thermometer is extensive.  The relative uncertainty in the energy per particle is thus 
\begin{equation}
\frac{\Delta\varepsilon}{\bar \varepsilon}= \frac{\Delta{E}}{\bar E}=\frac{1}{\sqrt{N}}\frac{\sqrt{\bar\varepsilon'}}{\bar \varepsilon}\,.\label{eqnsepsilon}
\end{equation}
Equation (\ref{eqnsepsilon}) and the identity $
\Delta \beta ={\Delta\varepsilon}/{\bar \varepsilon'}
$  together imply (\ref{eqn:relunc}), thus establishing the shot noise limit on extensive, thermalising thermometers.

This accords with the quantum Cram\'er-Rao bound in equation~(\ref{eq:quantumCramerBOUND}) on 
 temperature estimation, which depends on the system heat capacity $c_v$~\cite{zanardi_heatcapacity1, zanardi_heatcapacity2}
\begin{equation}\label{eq:CrboundT} 
  {\Delta T}^{\cal} \geq \frac{1}{ \sqrt{N {\cal Q} (T)}},  \quad {\cal Q} (T)= \frac{1}{k_B^2 T^4} \left(\mathrm{Tr} [\rho_\beta H^2] -\mathrm{Tr} [\rho_\beta H]^2 \,\right)=\frac{c_v}{k_B T^2},
\end{equation}
where ${\cal Q} (T)$ is the QFI~(\ref{def:QFI}) for the temperature $T$ and $1/\sqrt{N}$ is the SQL-scaling  in the number $N$ of independent probes. In other words, the ultimate limit to the precision at which the temperature of a thermal state can be determined is set by an energy measure on $\rho_\beta$. 
\footnote{It results that in this case the SLD commutes with the system Hamiltonian, implying that energy measurements are optimal. Moreover, thermal states represent a special case as they belong to the so-called exponential class~\cite{miller18}.}
The above inequality builds a first significant bridge between two apparently independent theoretical frameworks: 
 quantum thermodynamics and  quantum estimation theory. 

This scaling is potentially a significant issue for high precision thermometry. For instance, in Doppler gas thermometry, high precision spectroscopy of a gas in thermal equilibrium with a heat bath reveals the Maxwell-Boltzmann distribution of velocities \cite{PhysRevA.81.033848,PhysRevLett.98.250801,PhysRevLett.100.200801}, whose width is a direct measure of $k_B T$ of the gas. 
In recent Doppler thermometry experiments in alkali vapours the atomic flux through a beam of \mbox{$\sim10$} cm length and \mbox{$~2$ mm} diameter is $\dot N\sim 10^{15}$ atoms/sec \cite{truong:2015}.  The limit to the precision of such a thermometer is then $\Delta \beta \sim({\dot N\tau})^{-1/2}\approx10^{-7.5}\tau^{1/2}$, where $\tau$ is the integration time and $\Delta \beta \geq 1/\sqrt{N {\cal Q}(\beta)}$.  At 1 second this sets a maximum precision in measurements of $k_B$ of 1 part in $10^{7.5}$, which is about 1.5 orders of magnitude better than the current CODATA estimates for $k_B$.

\subsection{Heisenberg Limited Thermometry}

The SQL-like $1/\sqrt{N}$ scaling in  Eq.~(\ref{eqn:relunc}) is evocative of similar scaling in shot-noise limited phase estimation.  It is therefore interesting to question whether there is a way to improve this scaling in thermometry to the Heisenberg scaling limit, $\sim 1/N$.  This question was answered in the affirmative way in \cite{stace:2010a}, based on a constructive toy model which demonstrates this possibility.  

The thermometer in the toy model does not come to thermal equilibrium with the bath, and therefore is not subject to the argument that yields  Eq.~(\ref{eqn:relunc}).  Instead, the toy model takes advantage of several key observations:
\begin{enumerate}
\item Thermometry (like all physical measurements) can be cast as a counting problem.
\item Any counting measurement can be turned into a phase estimation problem.
\item Phase estimation can be performed with Heisenberg-limited scaling.
\end{enumerate} 
It follows that the thermometry can be done with an accuracy that improves as $1/N$.

To see how thermometry can be described as a counting problem, consider a bath of $M$ identical two-level atoms, each with energy splitting $\epsilon$ between the ground and excited state.  If the atoms are at some temperature $T$, the populations of the energy eigenstate  will be given by $\rho_\beta$.  This is plainly a thermodynamic quantity, so counting the number of excited atoms, $m$, is a proxy observable from which we can estimate the temperature via $\langle m\rangle =M\tfrac{e^{-\beta \epsilon}}{1+e^{-\beta \epsilon}}$.

\begin{figure}
\begin{center}
\includegraphics[width=8cm]{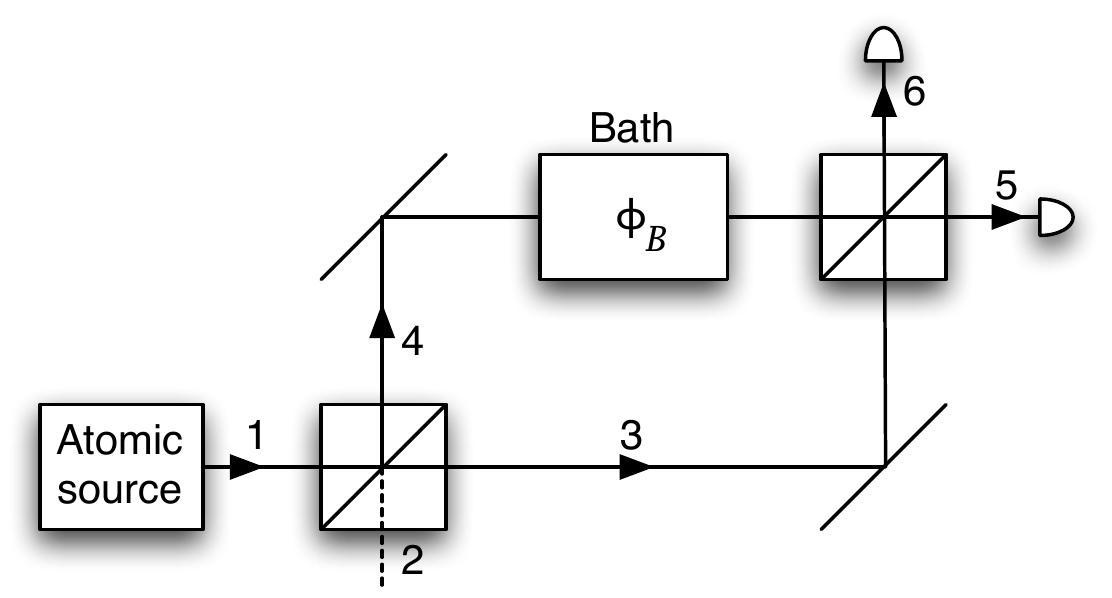}
\caption{An atomic interferometer with the bath in one branch.  Atoms from the thermometer are input into mode 1 and detected in modes 5 and 6.  Mode 2 is the vacuum port.  The figure is reprinted from  T. M. Stace, {\it Phys. Rev. A} {\bf 82}, 011611 (2010).} \label{fig2}
\end{center}
\end{figure}

To see how to turn a counting problem into a phase estimation problem, consider the following specially designed interaction Hamiltonian describing a dispersive coupling between the $M$ atoms in the thermal bath, and $N$ atoms which are part of a thermometric probe
\begin{equation}
H^{{\mathrm{(int)}}}=\alpha\sum_{j=1}^N\sum_{k=1}^M \ket{\epsilon}_j\bra{\epsilon}\otimes\ket{\epsilon}_k\bra{\epsilon}.\label{int}
\end{equation}
If the interaction between a given probe atom, prepared in its excited state $|{\epsilon}\rangle$ and the $M$ thermalised atoms lasts for  a time $\tau$, a phase $\phi_B=\alpha \,m\,\tau$ will accumulate on the probe atom.  Thus, counting $m$ can be turned into the problem of measuring the phase  $\phi_B$.  

This phase can be made observable if the probe atom can be introduced to an interferometer, with the interaction region in one branch of the interferometer, as shown in Fig.~\ref{fig2}.  Finally, as described earlier, if the atoms in the interferometer are prepared in a highly mode-entangled state (such as the well-known `NOON' state), it is possible to perform phase estimation with Heisenberg-limited scaling.

\subsection{Global versus Local Measurements}

What happens when technical or practical limitations restrict our capabilities to local probing? In such a case, the lower bound on the root-mean-square error on $T$ must be computed optimizing the Fisher Information over all possible local measurements on an accessible subsystem of $\cal S$. Without loss of generality, let us assume $\cal S$ to be composed of two subsystems $\cal A$ and $\cal B$, and to have access only to subsystem $\cal A$. In general, the global Hamiltonian shows local terms (acting on $\cal A$ or $\cal B$ separately) to be summed to interactions terms between the subsystems, i.e. $H=H_{\cal A} + H_{\cal B} + H_{\cal AB}^{\rm (int)}$. By definition, if we assume to perform a POVM on $\cal A$, the quantum Cram\'er-Rao bound~(\ref{eq:CrboundT})  reads
\begin{equation}\label{eq:CrboundT_A} 
  {\Delta T}^{\cal A} \geq 1/\sqrt{N {\cal Q}_{\cal A}(T)}= k_B T^2/ \sqrt{N \mathfrak{S}_{\cal A}[\rho_{\beta}]} \, ,
\end{equation}
where $\mathfrak{S}_{\cal A}$ is the so-called Local Quantum Thermal Susceptibility (LQTS) introduced with this name in~\cite{DePasquale16}, and corresponding to the QFI for the estimation of the inverse temperature $\beta$ computed on the reduced state $\rho^{\cal A}_\beta=\mathrm{Tr}_{\mathcal{B}}[\rho_\beta]$,
\begin{equation} \label{LQTS_def}
\mathfrak{S}_{\cal A}[\rho_{\beta}]:= 8 \lim_{\delta \beta \to 0} \frac{1-\mathcal{F} 
\big( \rho^{\cal A}_\beta, \rho^{\cal A}_{\beta+\delta \beta} \big) }{\delta \beta^2}\,. 
\end{equation}
This functional quantifies the ultimate precision limit to estimate the temperature $T$ 
by means of a local (quantum) measurement on subsystem ${\cal A}$, and by definition gauges how modifications on the global system temperature affect the local state $\rho^{\cal A}_\beta$: the larger is $\mathfrak{S}_{\cal A}[\rho_{\beta}]$ the more sensitive is the subsystem response. %
Notice that $H_{\cal AB}^{\rm (int)}$ can be arbitrarily strong.
A closed expression for LQTS can be determined by applying  Uhlmann's theorem for the fidelity~\cite{uhlmann76}, according to which the fidelity between the mixed states $\rho^{\cal A}_\beta$ and $\rho^{\cal A}_{\beta+\delta \beta}$ can be computed by means of a maximization over all possible purifications $|\rho_\beta\rangle$  and $|\rho_{\beta+\delta \beta}\rangle$ of such density matrices, through an ancillary system $a$ 
\footnote{Let $\rho^{\Gamma}$ be the state of a given system $\Gamma$. It is always possible to introduce another system $a$ (the so-called reference or ancillary system) and define a pure state $|\rho^{\Gamma}\rangle$ on $\Gamma a$ such that $\rho^{\Gamma} = \mathrm{Tr}_{a}[|\rho^{\Gamma}\rangle \langle \rho^{\Gamma}| ]$. Furthermore, if $|\rho^{\Gamma}\rangle$ and $|{\rho'}^{\Gamma}\rangle$ are two purifications of $\rho^{\Gamma}$ on $\Gamma a$ there exists a unitary transformation $U$ on $a$ such that $|{\rho'}^{\Gamma}\rangle = (\mathbb{I}_\Gamma \otimes U ) |\rho^{\Gamma}\rangle$, being $\mathbb{I}_\Gamma$ the identity operator on $\Gamma$. The last  property is known as ``freedom in purifications''~\cite{Nielsen_book}.}
\begin{equation}
  \label{eq:fidelity}
  \mathcal{F}\left(\rho^{\cal A}_\beta, \rho^{\cal A}_{\beta+\delta \beta}\right) 
  = \max_{|\rho_\beta\rangle,|\rho_{\beta+\delta \beta}\rangle} |\langle \rho_\beta 
  |\rho_{\beta+\delta \beta}\rangle|.
\end{equation} 
A convenient choice is to set the ancilla as $a = {\cal B A' B'}$, with ${\cal S' =\cal A'B'}$ isomorphic to ${\cal S =\cal AB}$, and $|\rho_\beta\rangle=\sum_i  (e^{- \beta E_i/2})/\sqrt{{\cal Z}_\beta} |E_i\rangle_{\cal A \cal B} \otimes |E_i\rangle_{\cal A' \cal B'}\;$,
being $H\!=\!\sum_i \!E_i |E_i\rangle_{\cal A \cal B} \langle E_i|$ the spectral decomposition 
of the system Hamiltonian. By exploiting the freedom in the purifications, it can be shown that the LQTS can be expressed as 
\begin{eqnarray}
  \label{eq:LQTSfinal}
\mathfrak{S}_{\cal A}[\rho_{\beta}]&=&\left(\mathrm{Tr} [\rho_\beta H^2] -\mathrm{Tr} [\rho_\beta H]^2 \,\right)- s_a
= \frac{c_v}{k_B\beta^2} - s_a
\end{eqnarray} 
with
\begin{eqnarray}
  \label{eq:sa}
s_{a} &=& \sum_{i<j} \frac{(e_i - e_j)^2}{e_i + e_j} |\langle e_i|H'| e_j \rangle |^2\,.
\end{eqnarray} 
Here $H'$ is the copy of $H$ acting on the isomorphic space, and $e_k$ and $|e_k \rangle$ are the eigenvalues and eigenvectors, respectively, of the ancilla density matrix $\rho_\beta^{a} = \mathrm{Tr}_{\cal A}[ |\rho_\beta\rangle \langle \rho_\beta|]$ sharing the same spectrum with $\rho_\beta^{\cal A} = \mathrm{Tr}_a[ |\rho_\beta\rangle \langle \rho_\beta|]$. It follows
\begin{equation}\label{eq:QFI_A}
{\cal Q}_{\cal A}(T)={\cal Q}(T) - \frac{s_{a}}{k_B^2 T^4}\,.   
\end{equation} 
Notice that  $s_a$ is always greater that zero, thus there is an ordering between the ultimate precision accuracy achievable via global and local measurements, ${\cal Q}_{\cal A}(T)\leq{\cal Q}(T)$, the inequality being saturated when ${\cal A}$ coincides with the whole system ${\cal S}$.
The same ordering holds at local level, since by construction, ${\cal Q}_{\cal A}(T)=\mathfrak{S}_{\cal A}[\rho_{\beta}]/(k_B^2 T^4)$ is a positive quantity which diminishes as the size of ${\cal A}$ is reduced, the smaller being the portion 
of the system we have access to, the worse being the accuracy we can achieve. 
\footnote{
The same conclusions can be driven by noticing an interesting connection between the LQTS, related to the reconstruction of $T$, and the more studied case of phase estimation mentioned in the former section. Indeed by comparing relations~(\ref{eq:QFI_nodisturb})-(\ref{eq:Qpure}) with~(\ref{eq:sa})-(\ref{eq:QFI_A}) we have
\begin{equation}
\mathfrak{S}_{\cal A}[\rho_{\beta}] + {\cal Q}_{a}(\lambda)= {\cal Q}_{\cal S S'}(\lambda) 
\end{equation}
where 
${\cal Q}_{a}(\lambda)$ and ${\cal Q}_{\cal S \cal S'}(\lambda)$ are the QFI associated to the estimation of $\lambda$ encoded on $|\rho_\beta\rangle$ via the unitary superoperator ${\cal E}_{\lambda/2}$ such that ${\cal E}_{\lambda/2}(|\rho_\beta\rangle \langle \rho_\beta| )=e^{-i H' \lambda/2}|\rho_\beta\rangle \langle \rho_\beta| e^{i H' \lambda/2}= |\rho_\beta^{(\lambda)}\rangle \langle \rho_\beta^{(\lambda)}|$, assuming to have access  at the measurement stage to subsystem $a={\cal B A' B'}$ and ${\cal S \cal S'} = {\cal A} a$, respectively:
\begin{equation}
{\cal Q}_{a}(\lambda)=\sum_{i<j} \frac{(e_i - e_{j})^2}{e_i + e_{j}}|\langle e_i | H' | e_j\rangle|^2\,, \quad
{\cal Q}_{\cal S S'}(\lambda)=\langle{\rho_\beta}| H'^2 | \rho_\beta \rangle - \langle{\rho_\beta}| H' |\rho_\beta \rangle^2 =\langle{\rho_\beta}| H^2 | \rho_\beta \rangle - \langle{\rho_\beta}| H |\rho_\beta \rangle^2.
\end{equation}
In other words the accuracies corresponding to the temperature estimation on $\cal A$ and the phase estimation on its complementary counterpart $a$, which are both positive quantities, are forced to sum up to the energy variance of the global system, thus establishing a sort of complementarity relation. Indeed, as already mentioned, the LQTS $\mathfrak{S}_{\cal A}[\rho_{\beta}]$ and similarly ${\cal Q}_{a}(\lambda)$ are increasing functions of the dimension of $\cal A$ and $a$, respectively.   
}

What is the role played by the interactions between the probed subsystem $\cal A$ and the remaining part $\cal B$ of the global system $\cal S$, the latter prepared in the thermal state $\rho_\beta$? In absence of interactions (i.e., $H_{\cal A B}^{\mathrm{(int)}} = 0$), the local state of $\cal A$ reads
\begin{equation}
\rho^{\cal A}_\beta= e^{- \beta H_{\cal A}}/{\cal Z_{\beta}^{\cal A}}
\end{equation}
 with ${\cal Z_{\beta}^{\cal A}}=\mathrm{Tr}[ e^{- \beta H_{\cal A}}]$, yielding  
\begin{equation}
 {\cal Q}_{\cal A}(T)=c^{\cal A}_v/(k_B T^2)\,,
 \end{equation}
 where $c^{\cal A}_v = \left(\mathrm{Tr} [\rho^{\cal A}_\beta H_{\cal A}^2] -\mathrm{Tr} [\rho^{\cal A}_\beta H_{\cal A}]^2 \,\right)/(k_B T^2)$ is the local heat-capacity of subsystem ${\cal A}$. Up to which limit does such local description hold in presence of interactions?
This problem has been explicitly tackled in the framework of locally interacting quantum systems~\cite{Gogolin16}, a very general class of models encompassing most of the fundamental spin models, such as the Ising model \cite{Ising25}, the Heisenberg model \cite{Bethe31}, the Potts model \cite{Wu82}, the Hubbard model \cite{Hubbard63}, etc (see Chapter~\ref{chap:19-Riera} for an extended review). A crucial property characterizing these Hamiltonians is that they admit a critical  temperature $T^*$ above which the correlation between any two observables $ O_{\cal A}$, $O'_{\cal A'}$ acting two subsystems $\cal A$ and $\cal A'$  of global system $\cal S$, the latter described by the thermal state $\rho_\beta$,
 decays exponentially with the distance $d(\cal A, \cal A')$ between their supports~\cite{Kliesch14}, i.e
\begin{equation}\label{eqcluster}
\Big|\mathrm{Tr}[\rho_\beta O_{\cal A}\; O'_{\cal A'}]-\mathrm{Tr}[\rho_\beta  O_{\cal A}] \mathrm{Tr}[\rho_\beta O'_{\cal A'}]\Big| \leq C_{\cal A A'}\left(\xi(T)+1\right) e^{-\frac{d({\cal A},{\cal A'})}{\xi(T)}}\,,
\end{equation}
where $C_{\cal A A'}$ is a constant fixed by $O_{\cal A}$ and $O'_{\cal A'}$ and $\xi(T)$ is the so-called correlation length of the system. 
It has been rigorously proved~\cite{DePalma17} that if the correlation length is much smaller than the volume to surface ratio of a given subsystem $\cal A$, the ${\cal Q}_{\cal A}(T)$ is a local quantity proportional to $c_v^{\cal A}$. Therefore, under this condition, local interactions between $\cal A$ and the remaining part of $\cal S$ (i.e. ${\cal B}$) do not significantly affect the precision of local measurements of the temperature. This condition is typically violated in the proximity of a critical point when the correlation length diverges, or for subsystems made by only few sites. 

\begin{figure}[h]
\begin{center}
{\includegraphics[width = 0.8 \textwidth]{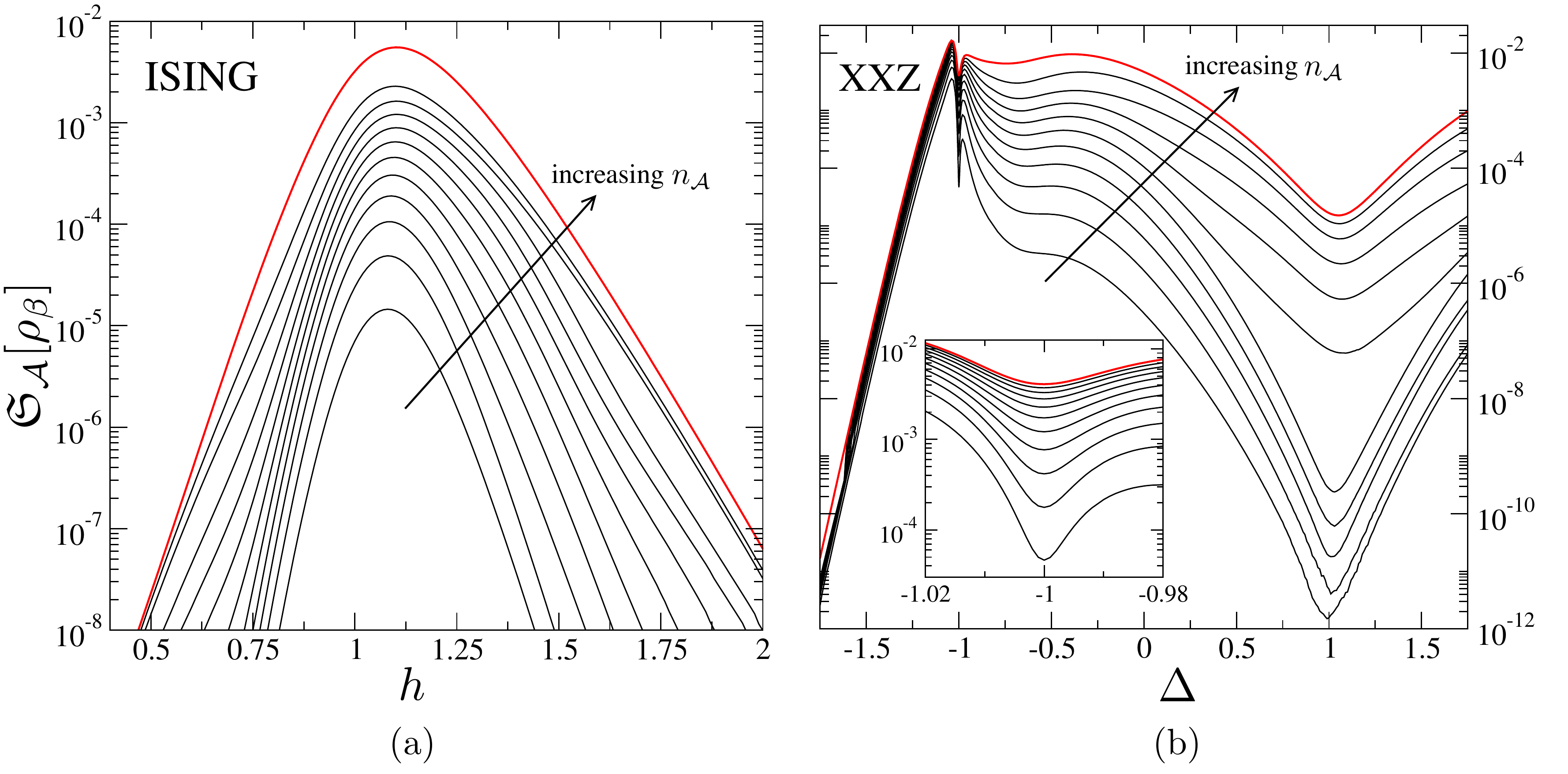}}
\caption{\label{fig:LQTSQPTpeak}  Numerical curves representing the behavior of the LQTS in the low-temperature regime ($\beta=9$) for the Ising (panel (a)) and the Heisenberg XXZ chains (panel (b)) with twelve sites. The uppermost  curve corresponds to the global quantum thermal susceptibility proportional to the heat capacity. The other curves have been computed  for different sizes $n_{\cal A}$ of the measured subsystem $\cal A$. In the XXZ model, the LQTS for $n_{\cal A}=1$ vanishes. The figure is reprinted from A. De Pasquale, et al., {\it Nat. Commun.} {\bf 7}, 12782 (2016), CC - by - 4.0 license.}
\end{center}
\end{figure}

\begin{figure}[h!]
\begin{center}
{\includegraphics[width = 0.8 \textwidth]{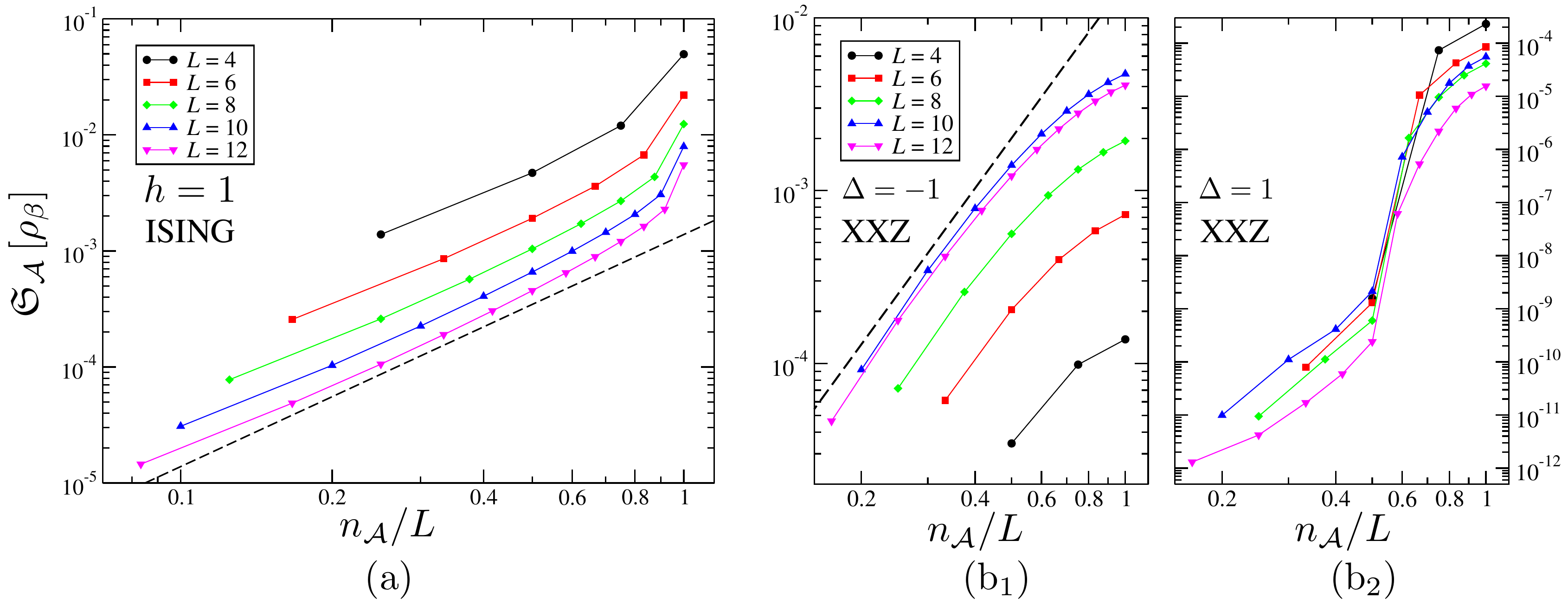}}
\caption{\label{fig:scaling} Analysis of the peak values of the LQTS for the Ising and the XXZ models, as a function of the dimension of the measured subsystem $n_{\cal A}$ and for different system lengths $L$.
   For the Ising model (panel (a)) the dashed line denotes a power-law behaviour $\mathfrak{S}_{\cal A} \sim (n_{\cal A}/L)^2$, 
    and is plotted as a guideline. For the XXZ chain (panels (b1) and (b2)) the data refer to the minima close to 
    the critical points $\Delta = \pm 1$. The dashed line for $\Delta=-1$ (panel (b1)) denotes the behaviour 
    $\mathfrak{S}_{\cal A} \sim (n_{\cal A}/L)^3$. In all panels we have set $h=1$ and $\beta = 3L/4$. The figure is reprinted from  A. De Pasquale, et. al., {\it Nat. Commun.} {\bf 7}, 12782 (2016), CC - by - 4.0 license.}
\end{center}
\end{figure}

Let us conclude this section by considering the thermal response at low temperature of two prototypical many-body systems featuring quantum phase transitions. In the specific, here we analyze the behavior of the quantum spin-$1/2$ Ising and Heisenberg chains, in a transverse magnetic field $h$ and with a $z$-axis anisotropy $\Delta$ respectively, both with periodic boundary conditions:
\begin{eqnarray}
  H_{\rm Ising} & \! = \! & 
  -  \sum_i \left[ \sigma_i^{x} \sigma_{i+1}^{x} + h \sigma_i^{z} \right] , \label{eq:Ising}\\
  H_{\rm XXZ} & \! = \! &  \sum_i \left[ (\sigma_i^{x} \sigma_{i+1}^{x} 
    + \sigma_i^{y} \sigma_{i+1}^{y}) 
    \! + \! \Delta \, \sigma_i^{z} \sigma_{i+1}^{z} \right] \!  \label{eq:XXZ}
\end{eqnarray}
(we have set to $1$ the system energy scale).
Here $\sigma^{\alpha}_i$ denotes the Pauli matrices on the $i$th site,  ($\alpha = x,y,z$). At zero temperature, the Ising model presents a $\mathbb{Z}_2$-symmetry breaking phase transition at $\vert h_{\rm c} \vert=1$, belonging to the Ising universality class. On the other hand, the XXZ-Hamiltonian shows a critical behaviour for $-1 \leq \Delta \leq 1$ and presents a ferromagnetic or antiferromagnetic ordering elsewhere, exhibiting in correspondence to the ferromagnetic point $\Delta = -1$ a first-order quantum phase transition, and a continuous one of the Kosterlitz-Thouless type
at the antiferromagnetic point $\Delta = 1$. Fig.~\ref{fig:LQTSQPTpeak} displays the small-temperature limit of the LQTS $\mathfrak{S}_{\cal A}[\rho_{\beta}]=  k_B^2 T^4 {\cal Q}_{\cal A} (T) ={\cal Q}_{\cal A}(T)/( k_B^2 \beta^4 )$ 
for these two Hamiltonians. Observe that as expected the LQTS is a monotonically 
increasing function of the number $n_{\cal A}$ of contiguous spins belonging to 
the tested subsystem ${\cal A}$. 
An interesting fact is that, even at finite temperatures and for systems composed of twelve sites, 
the LQTS seems to be sensitive to the presence of critical regions. This effect can be naively understood by observing that at low temperatures the Hamiltonian energy levels which play a significant role in the system dynamics are the ground state and the first excited levels, whose interplay underpins the emergence of quantum phase transitions. The sensitivity of LQTS to critical points can be understood by noticing that by definition it basically addresses the degree of distinguishability among such energy levels.
A fulfilling quantum-metrology approach to quantum phase transitions at finite temperatures can be found in~\cite{Zanardi07} and ~\cite{Mehboudi15}. 
The numerical analysis reported in Fig.~\ref{fig:scaling} of the scaling of the LQTS close to quantum phase transition points shows significant deviations from a linear growth with $n_{\cal A}$, indicating that, as expected, in this point the correlations cannot be neglected already for the sizes considered here.

\section{Single qubit thermometry}
The recent technological progress in manipulating individual quantum systems, has enabled their employment as temperature probes. Indeed, accurate temperature readings at nanoscopic level can  find applications in many research areas, ranging from materials science~\cite{Schwab00, Linden10} to medicine and biology~\cite{Klinkert09,Schirhagl14}, and in most of the situations addressed by the all field of quantum thermodynamics~\cite{Gemmer_book, Campisi11, Horodecki13, Carrega15, Carrega16} requiring the control of quantum thermal devices.
In this section we will tackle different aspects of a quite typical quantum estimation problem: the reconstruction of the unknown temperature of a sample by putting it in thermal contact with an individual quantum probe, acting as a thermometer.  

\subsection{Temperature Discrimination using a Single Qubit}

\begin{figure}[h!]
\begin{center}
\includegraphics[width=8cm]{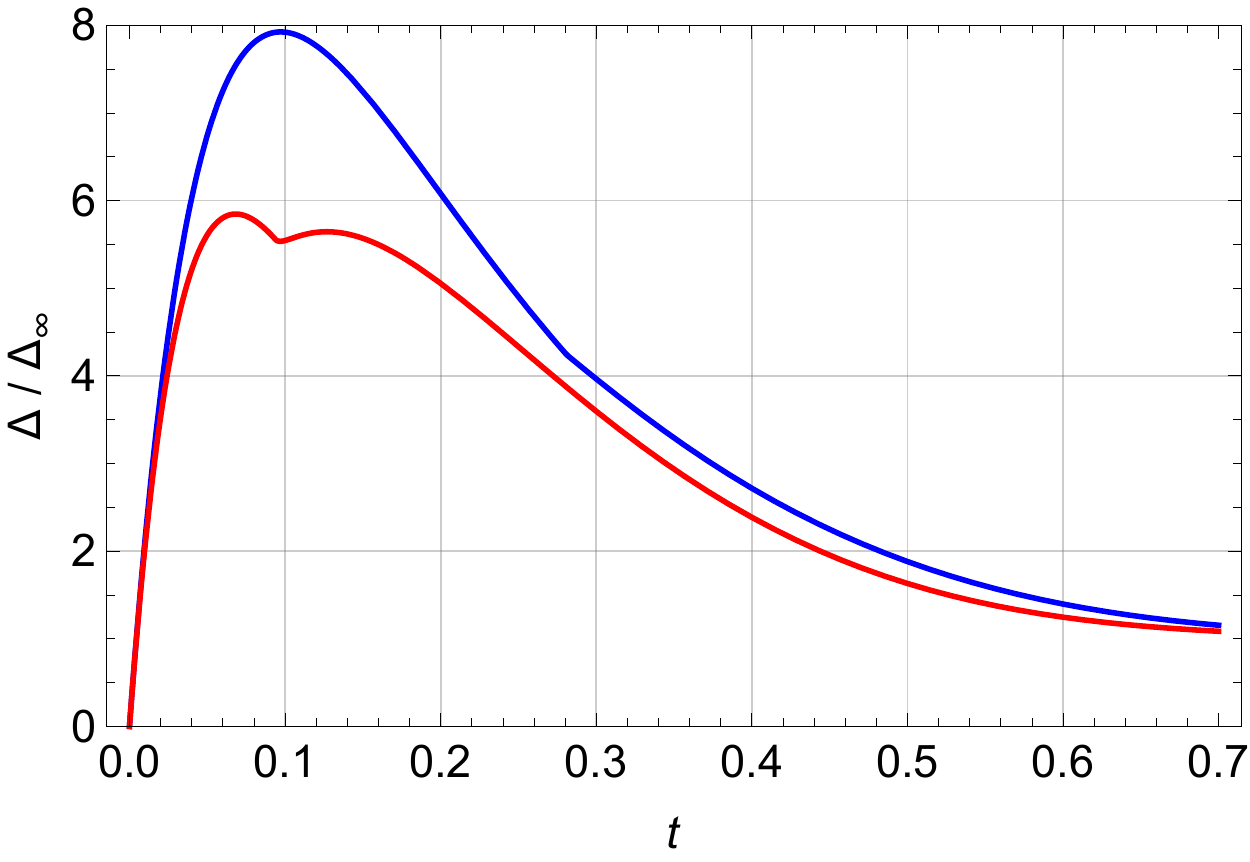}
\caption{Trace distance between the state of a probe qubit after thermalising with either a hot bath at temperature $T_h$, or a cold bath at temperature $T_c$. The distance measure $\Delta(\vec r_h(t), \vec r_c(t))$ is the distance between the Bloch vectors $\vec r_h(t)$ and $\vec r_c(t)$ of the corresponding qubit states, $\rho_h$ or $\rho_c$, after thermalisation for $t$ units of time.  The normalisation $\Delta_\infty$ is the distance at thermal equilibrium, i.e.\ at $t\rightarrow \infty$.  Figure is reprinted from S. Jevtic, et al., {\it Phys. Rev. A} {\bf 91}, 012331 (2015).} \label{fig3}
\end{center}
\end{figure}

Given a probe, plainly measuring the temperature is sufficient to discriminate between two temperatures.  However, the more elementary thermometric task of discriminating between whether a bath (a sample characterized by a large number of subcomponents) is `hot' and `cold' does not require a numerical temperature scale.  In the context of quantum thermometry, it was shown in \cite{jevtic:2015} that a quantum two-level system can perform this task.  We suppose the bath is at either $T_h$ or $T_c$, and we wish to optimally discriminate between these two.  A single qubit may be brought into thermal contact with the bath, and after interacting with it for some time $t$, the state of the qubit is measured.

In this scenario, the Fisher Information is not the relevant measure: we are not attempting to discriminate between a continuum of states, but between two distinct state of the qubit.  As such, the trace distance, which is given by the Euclidean distance $\Delta(\vec r_h(t), \vec r_c(t))$  between the Bloch vectors $\vec r_h(t)$ and $\vec r_c(t)$ of the possible states of the qubit, determines our ability to discriminate between the two states.  

Jevtic et al.\ \cite{jevtic:2015} studied this model as a function of the relative temperatures,  the interaction time, and the initial state of the probe qubit.  Assuming a Markovian interaction between the qubit and the bath (characterised by a thermalisation rate $\tilde{\gamma}$) they find that $\Delta(\vec r_h(t), \vec r_c(t))$ is maximised at a finite interaction time, and with the qubit initialised in the ground state. This is shown in the red curve in Fig. \ref{fig3}, where time is measured in units of $\tilde{\gamma}^{-1}$: up until the cusp around $t\approx0.1/\tilde{\gamma}$, the optimal initial state of the qubit is $\ket{g}$.  Clearly, the maximum value of that curve occurs in this interval, indicating that the globally optimal strategy is to initialise the qubit in the ground state, and run the interaction for $t\approx0.06/\tilde{\gamma}$.   If the interaction time is longer than $t\approx0.1/\tilde{\gamma}$,  the locally optimal initial state is no longer the ground state, but some coherent superposition of ground and excited.  Thus quantum coherence provides some enhancement in a limited sense in this temperature discrimination problem.

In \cite{jevtic:2015}, it was also shown that including an ancilla qubit, which is initialised in a maximally entangled state with the probe qubit yields better performance, as indicated by the blue curve in Fig.~\ref{fig3}.  Clearly it is higher than the red curve at all interaction times, indicating that quantum entanglement is a potentially useful resource in this state discrimination task.

\subsection{Fundamental limitations on temperature estimation with individual quantum probes}
We now discuss temperature reconstruction of a reservoir at thermal equilibrium. 
In standard thermometry, the thermometer is put in contact with the bath and then it is allowed to equilibrate, so that it finally ends up to be in a thermal state $\rho_{\beta}=e^{-\beta H}/{\cal Z}_\beta$. Therefore, as seen in the former section the highest achievable accuracy on temperature estimation through optimal measurements on the probe~(\ref{eq:CrboundT}) is proportional to the its heat capacity, that is to the variance of $H$:
\begin{equation} 
  \Delta T \geq \frac{1}{\sqrt{N \cal Q(T)}}, \quad {\cal Q} (T)= \frac{c_v}{k_B T^2}=\frac{1}{k_B^2 T^4} \left(\mathrm{Tr} [\rho_\beta H^2] -\mathrm{Tr} [\rho_\beta H]^2 \,\right)\,,
\end{equation}
where $N \gg 1$ is the number of probes at disposal, or the repetition of same the estimation procedure. 
It follows that the optimal choice of the probing system allowing for maximal thermal sensitivity, consists in finding the energy spectrum with the largest possible energy variance at thermal equilibrium. If, without loss of generality, the probing system is assumed to be $M$-dimensional, the solution to this problem amounts to solve $M$-coupled transcendental equations
\begin{equation}
\frac{\partial }{\partial E_m} \left[\frac{1}{{\cal Z}_\beta} \sum_i {E_i}^2 e^{-E_i/(k_B T)} -  \frac{1}{{{\cal Z}_\beta}^2} \left(\sum_i {E_i} e^{-E_i/(k_B T)}\right)^2\right]=0\,,
\end{equation}
where $H=\sum_i E_i |E_i \rangle \langle E_i |$ is the Hamiltonian spectral decomposition. It can be shown~\cite{Correa15, Reeb15} that the optimal quantum probes, acting as thermometers of maximal thermal sensitivity, are given by effective two-level atoms with maximally degenerate excited state and with a temperature dependent energy gap (this is equivalent to saying that for a fixed energy gap of the qubit thermometer there exists a temperature which can be retrieved with optimal accuracy).
\subsection{Single-qubit thermometry through sequential measurements.}
The analysis outlined above hinges upon two hypothesis:
\begin{enumerate}[label=\textit{\roman*)}]
\item The interaction time is sufficiently long to let the probe thermalize with the bath.
\item A certain number $N$ of probes prepared in the same input state, say $\rho_0$, interact with the bath and are measured independently, or equivalently if one has at disposal a single probe it is reinitialized in the same state after each of the $N$ measurement stages. In other words, the whole experiment requires $N$ independent and identically distributed (i.i.d.) measurements leading to the Cram\'er-Rao bound~(\ref{eq:CramerBOUND}).
\end{enumerate}
However, in practice $i)$ and $ii)$ can find some limitations. On the one hand, one may need to read the temperature of the outgoing probe before the latter attains full thermalization. This for instance happens if it is not possible to arrange the interaction time with the reservoir to be long enough, or if the bath itself is unstable (e.g. in the low temperature regime when too strong correlations are established between the probe and the sample~\cite{Correa16}, and fundamental limits emerge in temperature reconstruction~\cite{Hofer17}). In general if $i)$ is violated, but it is still possible to fulfill $ii)$, it has been observed~\cite{Correa15} that a convenient choice for $\rho_0$ is represented by the Hamiltonian ground state. On the other hand, also the arbitrary initialization of $N$ independent probing systems, or of the single probe at disposal after each measurement readout, might encounter some obstructions, thus violating condition $ii)$. In order to attack this limitation, one possibility is relying on sequential measurement schemes~\cite{Guta11,Burgarth15}, where repeated consecutive measurements are performed on a single probe without reinitializing it. For the sake of clarity let us indicate as ${\cal E}_T^{\tau}$ the superoperator defining the process that the probe undergoes when interacting with the sample, thus explicitly labelling it with the temperature $T$ of the sample, and with the interaction time $\tau$ between the probe and the bath. Furthermore, in order to provide a simple, but  mathematically rigorous, description of the sequential measurement protocol, it is useful to introduce a description also of the measurement process in terms of superoperators. In general, once a POVM, $\{\Pi_\theta\}, \int d \theta \Pi_\theta=\mathbb{I}$, has been selected, we can  associate to it a family of superoperators $\{{\cal M}_\theta\}$ fulfilling the normalization condition $\int d \theta {\cal M}_\theta={\cal I}$, with $\cal I$ being the identity superoperator  (i.e. ${\cal I}(\rho_0)=\rho_0$). When applied to an arbitrary state $\rho$, these superoperators give the outcome $\theta$ with probability $p(\theta|\rho)=\mathrm{Tr}[{\cal M}_\theta(\rho)]=\mathrm{Tr}[\Pi_\theta \rho]$, and transform $\rho$ as ${\cal M}_\theta(\rho)/\mathrm{Tr}[{\cal M}_\theta(\rho)]$.
\footnote{By definition the elements $\Pi_\theta$ of a POVM are positive operators. This implies that for each $\Pi_\theta$ there exists an other positive operator $M_\theta$, determined up to a unitary transformation, such that $\Pi_{\theta}=M_\theta^\dagger M_\theta$ and $\int d\theta M_\theta^\dagger M_\theta = \mathbb{I}$. Therefore the probability of measuring $\theta$ on a state $\rho$ is given by $p(\theta|\rho)=\mathrm{Tr}[\Pi_\theta \rho]= \mathrm{Tr}[ M_\theta \rho M_\theta^\dagger]$, and the normalized state of the system after the measurement reads $\rho_\theta=  M_\theta \rho M_\theta^\dagger/p(\theta|\rho)$. Finally, to each operator $M_\theta$ we can associate a superoperator ${\cal M}_\theta$ such that ${\cal M}_\theta(\rho)=M_\theta \rho M_\theta^\dagger$ and $\int d\theta {\cal M}_\theta = {\cal I}$, with ${\cal I}$ being the identity superoperator.
}

\begin{figure}[h]
	\begin{center}
	{\includegraphics[width = 1 \textwidth]{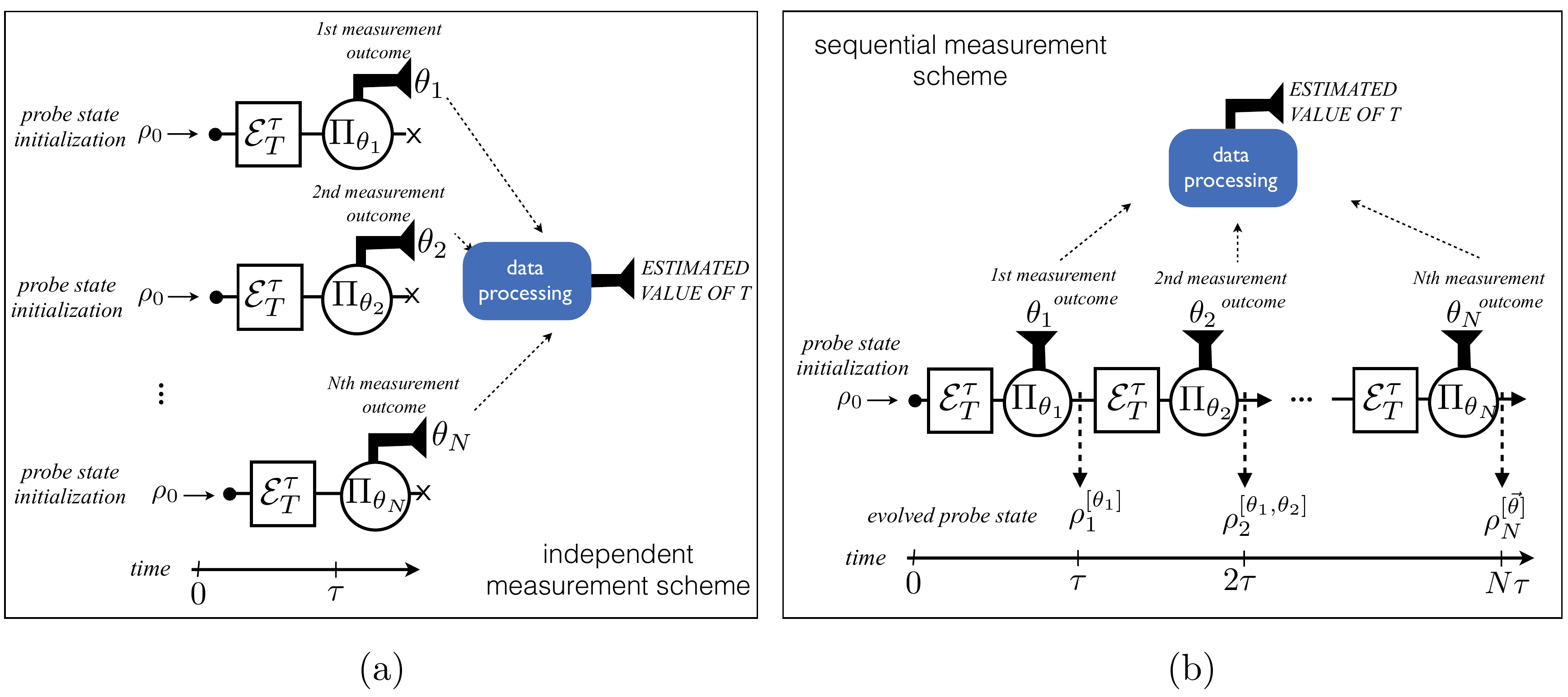}}
\caption{\label{fig:SMS} Schematic representation of two estimation temperature schemes realized by letting a probe (behaving as a thermometer) interact with a sample for a time interval $\tau$. Panel (a) represents the standard scheme relying on $N$ i.i.d. measurements performed on $N$ distinct probes (as in panel (a) of Fig.~\ref{fig:schemes}). Panel (b) refers to $N$ sequential measurements on the same probe which is initialized only once.}
\end{center}
\end{figure}

In Fig.~\ref{fig:SMS}  we have provided a pictorial representation of the two temperature estimation protocols we aim to compare.\\
\\
Let us stress that while in the standard approach the collected data $\vec{\theta}=\{\theta_1, \theta_2, \ldots \theta_N \}$ are independent identically distributed (i.i.d.), this is no more true for the sequential measurement scheme. Therefore, in the first case the global probability distribution of a $N$-long sequence $\vec{\theta}$ reads
\begin{equation} \label{NEWPROB}
 p^{(N)}_{\mathrm{i.i.d.}}(\vec{\theta}\,|T)= \prod_{i=1}^N p(\theta_i |T),
\end{equation} 
with $p(\theta_i|T)=\mathrm{Tr}[{\cal M}_{\theta_i} \circ {\cal E}_T^{\tau}(\rho_0)]$ all independent from each other, and the symbol ``$\circ$'' representing the composition of superoperators. On the contrary, in the second scenario we have
\begin{equation}
p^{(N)}_{\mathrm{s.m.s.}}(\vec{\theta}\,|T)
=\prod_{i=1}^N p(\theta_i |T;\theta_1,\ldots, \theta_{i-1}) \label{eq:p_seq1}
\end{equation} 
 with $p(\theta_i |T;\theta_1,\ldots, \theta_{i-1})=\mathrm{Tr}[({\cal M}_{\theta_i} \circ {\cal E}_T^{\tau})(\rho^{[\theta_1,\ldots, \theta_{i-1}]})]$, and 
 $\rho_0$, $\rho^{[\theta_1]}_1$, $\rho^{[\theta_1,\theta_2]}_2$, \ldots, $\rho^{[\theta_1,\ldots, \theta_N]}_N$ the density matrices generated by the measurement stochastic process
\begin{align}
\rho^{[\theta_1]}_1&= \frac{({\cal M}_{\theta_1}\circ {\cal E}_{T}^\tau)(\rho_0) }{\mathrm{Tr}[({\cal M}_{\theta_1}\circ {\cal E}_{T}^\tau )(\rho_0) ]}  &(t=\tau)  ,\nonumber\\
\rho^{[\theta_1,\theta_2]}_2&= \frac{({\cal M}_{\theta_2}\circ  {\cal E}_{T}^\tau)(\rho^{[\theta_1]}_1)}{\mathrm{Tr}[({\cal M}_{\theta_2}\circ  {\cal E}_{T}^\tau)(\rho^{[\theta_1]}_1)]}  &(t=2\tau), \nonumber\\
&\qquad\qquad\vdots&&\nonumber\\ 
\rho^{[\theta_1,\ldots, \theta_{N}]}_N&= \frac{({\cal M}_{\theta_N}\circ   {\cal E}_{T}^\tau)(\rho^{[\theta_1,\ldots, \theta_{N-1}]}_{N-1})}{\mathrm{Tr}[({\cal M}_{\theta_N}\circ  {\cal E}_{T}^\tau  )(\rho^{[\theta_1,\ldots, \theta_{N-1}]}_{N-1})]} & (t=N\tau)\,.
\end{align}
Here, we have also assumed to neglect the measurement time. Notice that $p^{(N)}_{\mathrm{s.m.s.}}(\vec{\theta}\,|T)$ can be equivalently written as 
$p^{(N)}_{\mathrm{s.m.s.}}(\vec{\theta}\,|T)= \mathrm{Tr}[({\cal M}_{\theta_N} \circ {\cal E}_{T}^\tau \circ {\cal M}_{\theta_{N-1}}  \circ {\cal E}_{T}^\tau
 {}\circ\cdots \circ {\cal M}_{\theta_1} \circ {\cal E}_{T}^\tau) (\rho_0)]$.

 The difference between the two approaches clearly emerges in the computation of the Cram\'er-Rao bound. 
For the case of i.i.d. readout, we get the scaling in~(\ref{eq:CramerBOUND}) given by $1/\sqrt{N}$
\begin{equation}\label{CR}
\Delta T_{\mathrm{i.i.d.}}^{(N)} \geq \frac{1}{\sqrt{\mathcal{F}_{\mathrm{i.i.d.}}^{(N)}(T)}} = \frac{1}{\sqrt{ N \mathcal{F}(T)}}\,,
\end{equation} 
where ${\cal F}(T)$ is the Fisher Information associated to the selected POVM, i.e. 
\begin{equation}\label{eq:F_nIID}
 \mathcal{F}^{(N)}_{\mathrm{i.i.d.}} = \int d\vec{\theta} \frac{1}{p^{(N)}_{\mathrm{i.i.d.}}(\vec{\theta}\,|T)} \left(\frac{\partial  p^{(N)}_{\mathrm{i.i.d.}}(\vec{\theta}\,|T)}{\partial T}\right)^2  
  = N   \mathcal{F}(T), \qquad \mathcal{F}(T)=\int\,d\theta \frac{1}{p(\theta|T)} \left(\frac{\partial p(\theta|T)}{\partial T}\right)^2\,.
\end{equation}
The same scaling does not hold for sequential measurements since the integral over $d \vec{\theta}$ does not factorise, yielding
\begin{equation} \label{CRSMS}
\Delta T^{(N)}_{\mathrm{s.m.s.}}\geq \frac{1}{\sqrt{  \mathcal{F}^{(N)}_{\mathrm{s.m.s.}}(T)}}, \quad  \mathcal{F}^{(N)}_{\mathrm{s.m.s}}(T) =   \int d \vec{\theta} \frac{1}{p^{(N)}_{\mathrm{s.m.s.}}(\vec{\theta} | T)} \left(\frac{\partial  p^{(N)}_{\mathrm{s.m.s.}}(\vec{\theta} | T)}{\partial T}\right)^2 \,.
\end{equation}
Finally, the quantum Cram\'er-Rao bound can be determined via a maximization over all possible POVMs. 

\begin{figure}[h]
	\begin{center}
	{\includegraphics[width = 0.8 \textwidth]{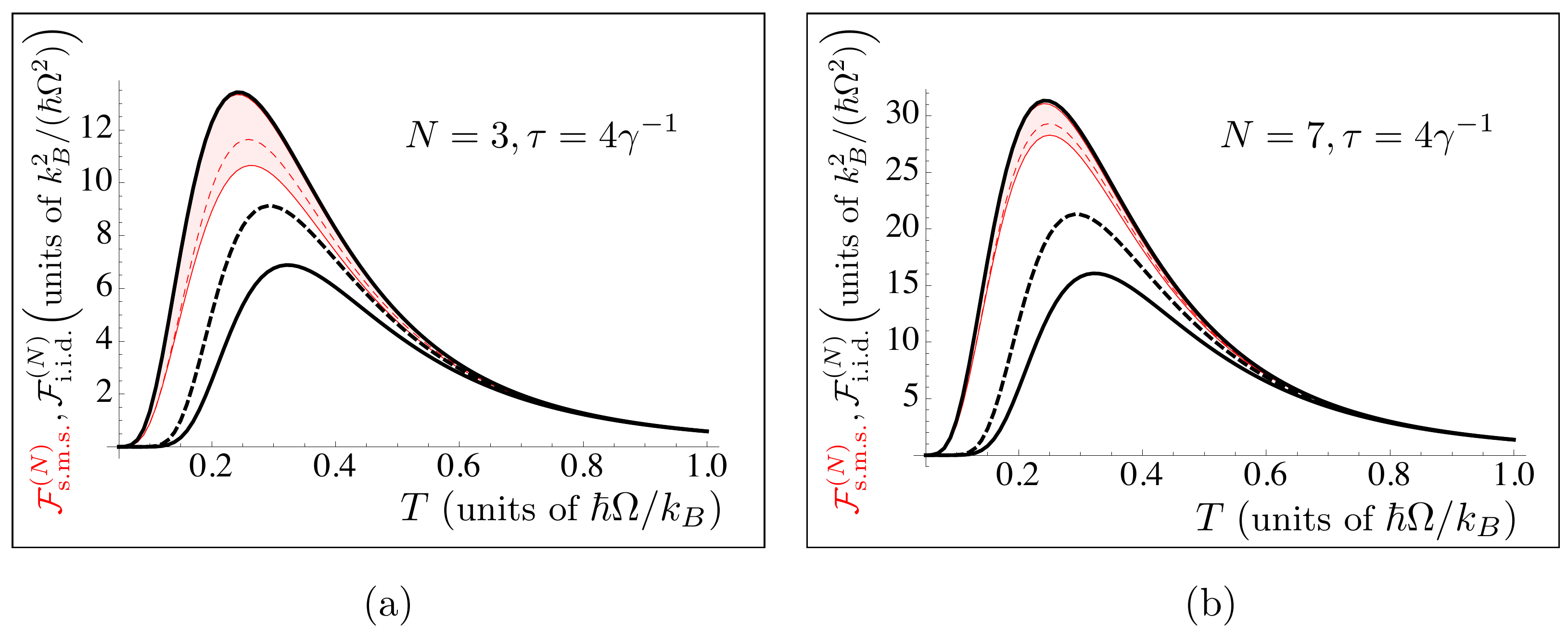}}
\caption{\label{fig:FI_iid_vs_sms} Fisher Information for projective measurements on the probe. The panels refer to different values of the number of repetitions $N$ and to an interaction time $\tau= 4 \gamma^{-1}$. Thick (black) lines refer to the standard protocol based on i.i.d. measurements while the thin (red) ones refer to sequential measurements (for the latter case the region between the optimal and worst performance ruled by the choice of the initial state of probe is shaded). The figure is reprinted from  A. De Pasquale, et al., {\it Phys. Rev. A} {\bf 96}, 012316 (2017).}
\end{center}
\end{figure}

\begin{figure}[h]
\includegraphics[width=0.4\columnwidth]{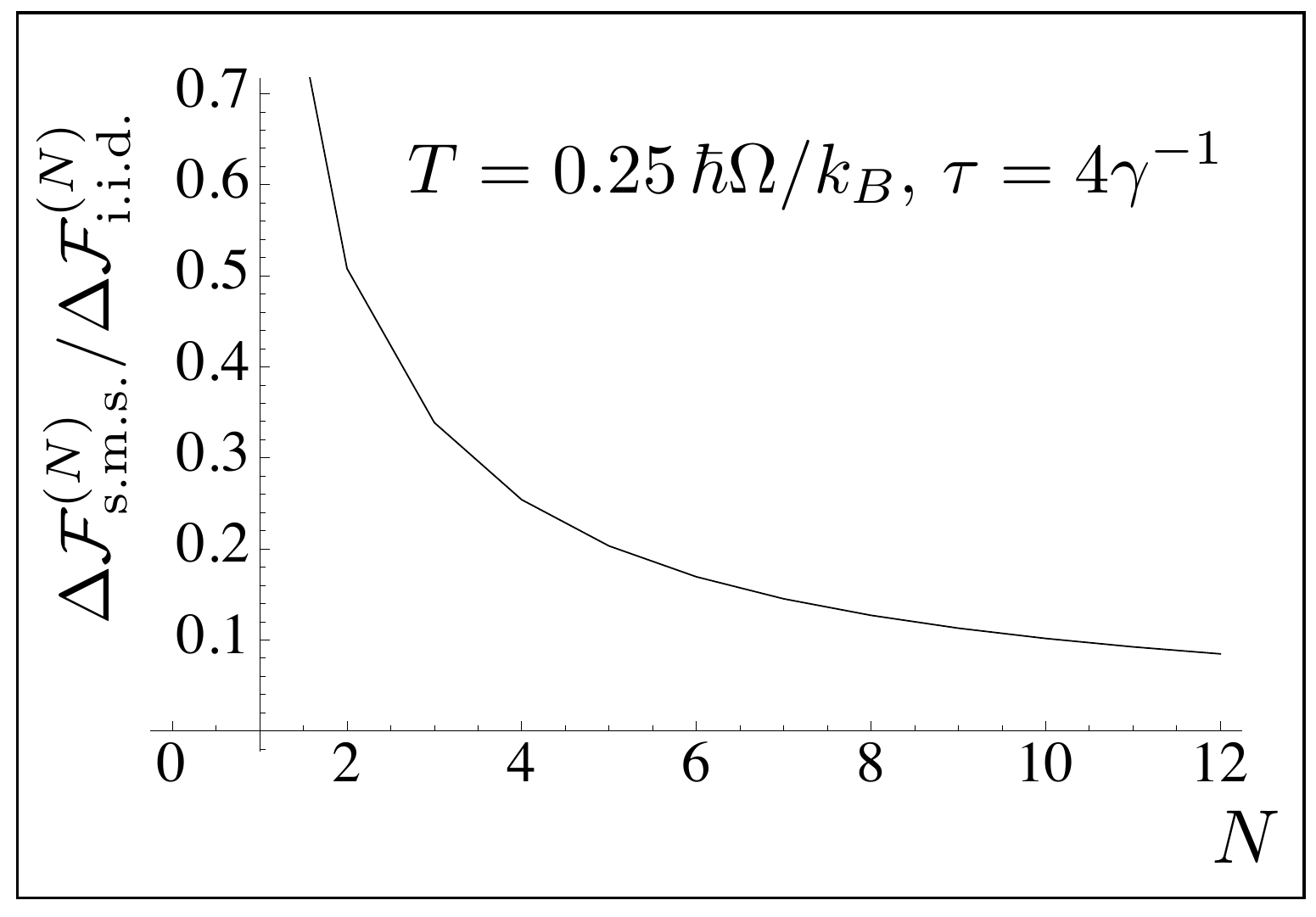}
\caption{Plot of the ratio between $\Delta\mathcal{F}^{(N)}_\mathrm{s.m.s}$ and $\Delta\mathcal{F}^{(N)}_\mathrm{i.i.d}$, the max-min gap of the FI due to the choice of the initial state for the probe. It results that $\Delta\mathcal{F}^{(N)}_\mathrm{s.m.s}$ shrinks  more rapidly then $\Delta\mathcal{F}^{(N)}_\mathrm{i.i.d.}$.  The figure is reprinted from  A. De Pasquale, et al., {\it Phys. Rev. A} {\bf 96}, 012316 (2017).}\label{fig:gap}
\end{figure}

Actually, in what follows we will assume a more practical perspective and represent some results~\cite{DePasquale17} dealing with a quite standard model for the reservoir and for the probe-sample interaction, and referring to a specific choice for the readout measurements. In the specific we consider a qubit system initialised in the state $\rho_0$ and put in contact, at time $t=0$, with a Bosonic  thermal reservoir of unknown temperature $T$. At $t>0$ the state of probe reads  $\rho_T(t)={\cal E}_T^t (\rho_0)=  e^{t \mathcal{L}_T}(\rho_0)$, where $\mathcal{L}_T$ is the Lindblad  superoperator defined as
\begin{equation}\label{eq:master} 
\mathcal{L}_T({\cdots})
= {-\frac{i}{2}} \Omega [\sigma_z, ({\cdots})]_-   +  \sum_{s= \pm1} \gamma_{s}    \left(\sigma_{-s} ({\cdots}) \sigma_{s} -  \frac{1}{2}[ \sigma_{s}\sigma_{-s}, ({\cdots})]_+ \right)  ,
\end{equation} 
with $[({\cdots}),({\cdots})]_\pm$ being the commutator ($-$) and anti-commutator ($+$) brackets, and $\sigma_{\pm}=(\sigma_x \pm i \sigma_y)/2$ being the spin-flip operators. In this expression,  the two relaxation constants  $\gamma_{-}$ (for excitation) and    $\gamma_{+}$ (for decay) are related to the temperature $T$ of the reservoir through the detailed balance condition and are given by
\begin{equation}
\gamma_{+} = (1+N_\mathrm{th}) \gamma, \qquad \gamma_{-}=N_\mathrm{th} \gamma\,,
\end{equation} 
where $N_\mathrm{th}=1/(e^{\beta\hbar\Omega}-1)$ is the average thermal number of Bosonic bath excitations, and $\gamma$ is a temperature-independent parameter gauging the strength of the probe-sample interactions. The qubit acts a thermometer, and $T$ is recovered by monitoring the populations of its two energy levels at time intervals $\tau$. The simplest POVM which can be realized to this end is given by the rank-one projectors $\Pi_{\pm} =(\mathbb{I} \pm \sigma_z)/2$ on the energy levels of the probe Hamiltonian $\frac{1}{2}\hbar\Omega\sigma_z$. In Fig.~\ref{fig:FI_iid_vs_sms} the FI associated to the two above mentioned estimation schemes,~(\ref{eq:F_nIID}) and~(\ref{CRSMS}), are plotted by setting the number measurements to be equal to $3$ and $7$, and the probe-sample interaction time as $\tau=4 \gamma^{-1}$ (the full thermalization time of the probes can be considered reached already for $\tau \gtrsim 9.5 \gamma^{-1}$). The uppermost and lowest solid lines correspond to the optimal and worst choices of the input state $\rho_0$ (which coincide with the ground state and with the first excited level, respectively) while the dashed lines refer to the average of the FI over uniformly  sampled input probe states $\rho_0$ (physically the latter curves correspond to the practical situation in which the experimenter is not able to completely control the probe preparation stage). It results that for the optimal choice of the probe input state the standard protocol based on i.i.d. measurements slightly outperforms the sequential measurement scheme. However a more interesting phenomenon is observed for non-optimal input states: for all temperatures of the bath, the sequential measurement protocol outperforms the standard one. Furthermore, also the gap between 
the optimal and the worst choice of the probe is smaller for sequential measurements than for i.i.d. ones, thus yielding a higher degree of versatility with respect to the choice of the initial state of the probe. This effect can be explained by observing that in this estimation approach the probe  gradually loses the memory of the initial condition, moving towards a fixed-point configuration depending on the bath temperature. It results that even a non-optimal initialization of the probe can in the end provide a relatively good estimation of the temperature. Fig.~\ref{fig:gap} shows that the gap between optimal and worst performances in case of sequential measurements $\Delta\mathcal{F}^{(N)}_{\mathrm{s.m.s.}}$ is lower and closes faster than that yielded in standard case indicated as $\Delta\mathcal{F}^{(N)}_{\mathrm{i.i.d}}$. Finally, when the probe thermalization with the bath has taken place, it can be shown that all the above mentioned curves collapse. 
 
\section{Conclusions}

We have discussed the application of techniques from quantum estimation theory and quantum metrology to thermometry.

It results that the Quantum Fisher Information plays an important role, and we have presented a toy model which demonstrates the possibility of Heisenberg-like scaling for thermometry.  
Then, we have introduced a theoretical approach to temperature locality, aiming at avoiding any restrictive hypothesis on the dimension of the system or on its Hamiltonian ruling the interactions between its subcomponents. This has led to the definition of the so-called local quantum thermal susceptibility, a functional which operationally highlights the degree at which the thermal equilibrium of the global system is perceived locally, and reduces to the system heat capacity when the global system is probed. Finally, moving from the observation that two-level quantum systems can be exploited as efficient thermometers to establish the temperature of a thermal bath, we have discussed different techniques of single qubit thermometry.

\bigskip

\acknowledgements 
\section{Acknowledgements}
ADP acknowledges financial support from the University of Florence in the framework of the University Strategic Project
Program 2015 (project BRS00215).

\end{document}